\documentstyle[prl,aps,epsfig,citesort]{revtex}

\newcommand{\beq}{\begin{equation}}
\newcommand{\eeq}{\end{equation}}
\newcommand{\beqas}{\begin{eqnarray*}}
\newcommand{\eeqas}{\end{eqnarray*}}
\renewcommand{\vec}[1]{\mbox{\boldmath$#1$}}

\begin{document}
\twocolumn[\hsize\textwidth\columnwidth\hsize\csname 
@twocolumnfalse\endcsname

\title{Two-dimensional Granular Gas of Inelastic 
       Spheres with Multiplicative Driving}

\author{Raffaele Cafiero (1), Stefan Luding (2), 
        and Hans J\"urgen Herrmann (1,2)}
\address{
(1) P.M.M.H., Ecole Sup\`erieure de Physique et de Chimie 
 Industrielles (ESPCI), \\ 
 10, rue Vauquelin, 75251 Paris CEDEX 05, FRANCE \\
(2) Institute for Computer Applications 1, 
 Pfaffenwaldring 27, 70569 Stuttgart, GERMANY\\}

\maketitle

\begin{abstract}
We study a two-dimensional granular gas of inelastic spheres subject to  
multiplicative driving proportional to a power $|v(\vec{x})|^{\delta}$ of 
the local particle velocity $v(\vec{x})$.  
The steady state properties of the model are examined for 
different values of $\delta$, and compared with the homogeneous 
case $\delta=0$. A driving linearly 
proportional to $v(\vec{x})$ seems to reproduce some 
experimental observations which could not 
be reproduced by a homogeneous driving.  
Furthermore, we obtain that the system can be homogenized
even for strong dissipation, if a driving inversely proportional to
the velocity is used.
\end{abstract}

\narrowtext
\vskip2pc]


Granular materials are a fascinating non-linear, dissipative,
non-equilibrium system \cite{herrmann98,haff83}, which are of
interest, first due to their practical importance and second because 
of the challenges they represent theoretically and numerically, but
experimentally as well. Granular media are collections of macroscopic 
particles with rough surfaces and dissipative, frictional interactions. 
For the sake of simplicity, we restrict ourselves to the smooth
situation here, and use the inelastic hard sphere (IHS) model
\cite{goldhirsch93,luding98d,n98,noije98}.  For more realistic 
interaction models see Refs.\ \cite{herrmann98,cafiero99} and 
the references therein.

One of the outstanding effects is the so called clustering,
a self-stabilized density instability due to dissipation 
\cite{goldhirsch93,luding98d} where
large, dense collections of particles coexist with almost
empty areas.  Clustering occurs in initially homogeneous systems  
\cite{haff83,goldhirsch93,huthmann98,luding98d}
and should not be confused with the so-called ``inelastic collapse'' 
\cite{goldhirsch93}, the divergence of the collision rate, 
which is inherent to the frequently used hard (rigid) sphere model 
\cite{goldhirsch93,luding98d,cafiero99}.
Freely cooling systems -- mainly examined numerically and 
theoretically \cite{goldhirsch93} -- are almost
impossible to realize experimentally.  Only recently, laboratory experiments 
were performed, where clustering could be examined in driven systems
\cite{kudrolli97,olafsen98,losert98},
where also kinetic theory approaches complemented by numerical 
simulations have proven to be successful
\cite{duparcmeur95,puglisi98,bizon99,n98,noije98,cafiero99,nie99}.

The driving of a granular material can be realized by moving walls 
\cite{herrmann98} which lead to local heating, or the system can 
alternatively be driven by a global homogeneous, random energy source 
in different variations \cite{puglisi98,bizon99,n98,noije98}. 
The latter type of energy input does not exactly resemble
the experimental situation, where a two-dimensional (2D) horizontal layer
of spheres is agitated by vertical vibrations of the bottom surface
and the horizontal degrees of freedom are indirectly agitated due to
the different vertical jump heights of the colliding particles 
\cite{olafsen98,nie99}. Thus, the choice of the driving term to 
put into a theory for dissipative systems is an open problem and 
we expect that it depends on the nature of the driving 
(vibrating wall, airflow, Brownian noise, etc.).  
In order to find possible candidates for a realistic energy input,
the IHS model with an inhomogeneous, {\em multiplicative driving} is
examined in the following.  The driving is proportional to the local velocity 
$|v(\vec{x})|^{\delta}$, with a given power $\delta$ which can take both
positive and negative values. The classical homogeneous driving with 
$\delta=0$ is contained in our approach as a special case. A positive
power leads to weak energy input for slow particles, e.g.~those moving
collectively inside a cluster \cite{olafsen98,nie99}.

In this letter a system of $N$ three-dimensional spheres with
radius $a$ and mass $m$ is considered,
interacting via a hard-core potential and confined to a 2D
plane of linear extension $L$, with periodic boundary conditions. 
The degrees of freedom are the positions $\vec r_{i}(t)$
and the translational velocities $\vec v_{i}(t)$ for each
sphere numbered by $i=1, \ldots, N$. The dissipation at a collision
is quantified by a constant normal restitution $r$.
From the momentum conservation law we can derive the change of 
linear momentum $-(m/2) (1+r) \vec{v}_{\rm c}^{(n)}$ of 
particle $i$ which collides with particle $j$,
with the normal relative velocity 
$\vec{v}_{\rm c}^{(n)} = [(\vec{v}_i-\vec{v}_j) \cdot \vec{n} ] \vec{n}$,
and the unit vector in normal direction
$\vec n = (\vec r_i - \vec r_j) / | \vec r_i - \vec r_j |$. 
The system is agitated each time interval $\Delta t=f_{\rm dr}^{-1}$, 
with a driving rate $f_{\rm dr}$ according to the frequency of the 
bottom. For homogeneous driving, it is costumary to assume 
$f_{\rm dr}\gg \omega$, where $\omega$ is the collision 
frequency of the granular gas \cite{n98,noije98}. This, however, 
does not correspond to the experiments, were $f_{\rm dr}$ 
can be rather small -- we will use driving frequencies around 
$100$\,s$^{-1}$, comparable to experimental situations \cite{olafsen98}. 
Numerical checks with strongly different values of $f_{\rm dr}$ 
lead to a similar behavior of the system even for driving frequencies 
lower than, but of the same order as $\omega$, provided that a stationary 
state is reached.  The velocity of particle $i$ is modified at
each time of agitation $t$ so that
\begin{eqnarray}
v'^x _i(t) & = & v_i^x(t)+r_i^x |\vec{v}_i(t)|^{\delta} v_r^{1-\delta} \nonumber \\
v'^y _i(t) & = & v_i^y(t)+r_i^y |\vec{v}_i(t)|^{\delta} v_r^{1-\delta} ~,
\label{eq:vxy}
\end{eqnarray}
where the prime on the left hand side indicates the value after 
the driving event. $v_r$ is
a reference velocity (in this study we use $v_r=1$\,m\,s$^{-1}$)
which allows to define the dimensionless translational particle
temperature $T=E/(N T_r)$, with
$E = (m/2) \sum_{i=1}^N $\mbox{\boldmath$v$}$_i^2$ 
and the reference temperature $T_r=m v_r^2$.
The variance of the uncorrelated Gaussian random numbers 
$r_i^x$ and $r_i^y$ (with zero mean) can now be interpreted 
as a dimensionless driving temperature $T_{\rm dr}$.  
The stochastic driving rule in Eq.\ (\ref{eq:vxy}) leads 
thus to an average rate of change of temperature 
\begin{equation}
\Delta T/\Delta t = H_{\rm dr} T^\delta 
{\rm ~~ with  ~~} H_{\rm dr} = f_{\rm dr} T_{\rm dr} ~.
\label{eq:Hdr}
\end{equation}
The time evolution equation of $T$
was derived for the case of a freely cooling granular gas
by means of a pseudo-Liouville operator formalism 
\cite{huthmann98,luding98d}. We adopt their nomenclature and 
account for the driving by adding Eq.\ (\ref{eq:Hdr}) to 
the mean field (MF) equation for the translational degree of freedom 
\begin{equation}
  \frac{d}{dt}  T_{}(t) = - G_r A T_{}^{3/2} + H_{\rm dr} T^\delta~.
\label{eq:mfrpT}
\end{equation}
For our case of a homogeneous monolayer of smooth spheres, one has 
$G_r  = 8 a n \sqrt{{ \pi T_r}/{m}} g(\nu) 
$, and $A = \frac{1-r^2}{4}$,
with the number density $n=N/V$, the pair correlation function
at contact $g(\nu)$, and the area fraction $\nu = N \pi a^2 / V$
covered by particles \cite{huthmann98,luding98d,cafiero99}. 
For the two-dimensional layer of spheres we
use the approximation
$g(\nu)=(1-7\nu/16)/(1-\nu)^2$ \cite{jenkins85b}.
For $\delta = 0$ the driving is homogeneous and independent of the
local granular temperature (or velocity). In the case $\delta\ne 0$
the driving is a function of $T_{}$ \cite{footnote1}.
Imposing
$\frac{d}{dt}  T_{}(t)=0$ one gets from Eq.\ (\ref{eq:mfrpT})
the MF temperature in the steady state
\begin{equation}
 T_{}^{\rm mf} = \left ( \frac{ H_{\rm dr} }
                        { G_r A } \right )^{\frac{2}{3-2 \delta}} ~,
\label{eq:mfeq}
\end{equation}
the generalization of the Enskog equilibrium
solution for a driven granular gas \cite{n98,noije98}. 
The scaling exponent of $T_{}^{\rm mf}$ in Eq.\ \ref{eq:mfeq} is $2/3$ 
for $\delta=0$, while it is $2$ for $\delta=1$. For $\delta\geq3/2$ the
 MF theory does not admit a stable equilibrium state 
\cite{cafiero99,footnote2}.

The final approach to the steady state can be obtained by linearizing
Eq.\ (\ref{eq:mfrpT}) around $T_{}^{\rm mf}$, what leads to an exponentially
fast approach to equilibrium with the power
${-[3 A \omega + \delta H_{\rm dr}(T_{}^{\rm mf})^{\delta-1}] t}$,
where $\omega=G_r \sqrt{T_{}^{\rm mf}} / 2$ is the Enskog collision frequency
for elastic particles with temperature $T_{}^{\rm mf}$. 
By inserting Eq.\ (\ref{eq:mfeq}) in the expression for $\omega$,
one can express the characteristic relaxation time
$t_{\rm relax}=[ \ldots ]^{-1}$ of $T_{}(t)$ 
as a function of the model parameters,
which reduces for $\delta=1$ to $t_{\rm relax}^{-1} = (5/2) { H_{\rm dr}}$.
Thus, for $\delta=1$ the characteristic time for the evolution of
$T_{}$ towards its equilibrium value does {\em not} depend on $A$ which
contains all the information about the inelasticity. This characteristics
is confirmed by numerical simulations.

Most of our event driven (ED) molecular dynamics simulations,
see \cite{huthmann98,luding98d,cafiero99} for details, 
with the driving specified in Eq.\ (\ref{eq:vxy}), are first 
equilibrated without driving and with elastic interactions ($r=1$),
until the velocity distribution is close to a Maxwellian. Then,
dissipation and driving are switched on. However, we checked that 
the steady state does not depend on the initial conditions.
The simulations are performed at fixed volume fraction
$\nu=0.34$ with $N=1089$ ($f_{\rm dr}=133$\,s$^{-1}$) or 
$N=11025$ ($f_{\rm dr}=67$\,s$^{-1}$), and different values of
$r$.  In our simulations we have chosen $a=10^{-3}$\,m and 
$H_{\rm dr}=1.0$\,s$^{-1}$ so that, for example, 
$G_r=8 \nu v_r/ \sqrt{\pi} a g(\nu) = 3.1 \times 10^{3}$\,s$^{-1}$,
$T^{\rm mf}=0.0358$ and thus $\omega=2.9 \times 10^{2}$\,s$^{-1}$,
if $r=0.90$ is used.
With these typical values and a homogeneous driving
($\delta=0$), the model of a driven granular gas is very close to a 
homogeneous state; no clusters are observed and the velocity distribution 
is almost Maxwellian \cite{cafiero99}.  This is in contradiction to
the experimental findings \cite{olafsen98}, and
suggests that the correct representation of the driving in those
experiments is {\em not the homogeneous white noise} usually implemented.
\begin{figure}[htb]
~\vspace{-1.9cm}\\
\begin{center}
 \epsfig{file=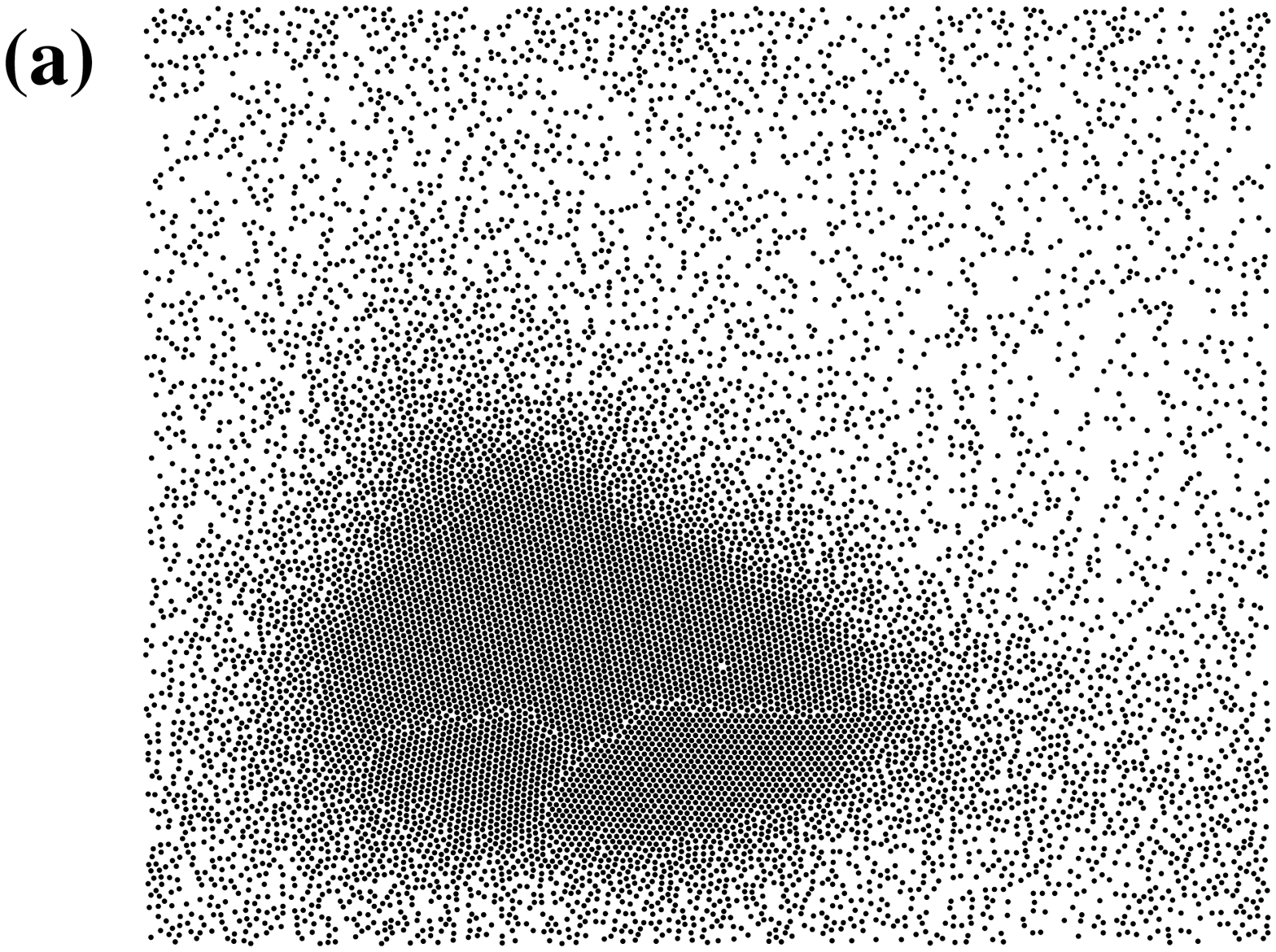,height=5.0cm,angle=-90} \\ 
 \epsfig{file=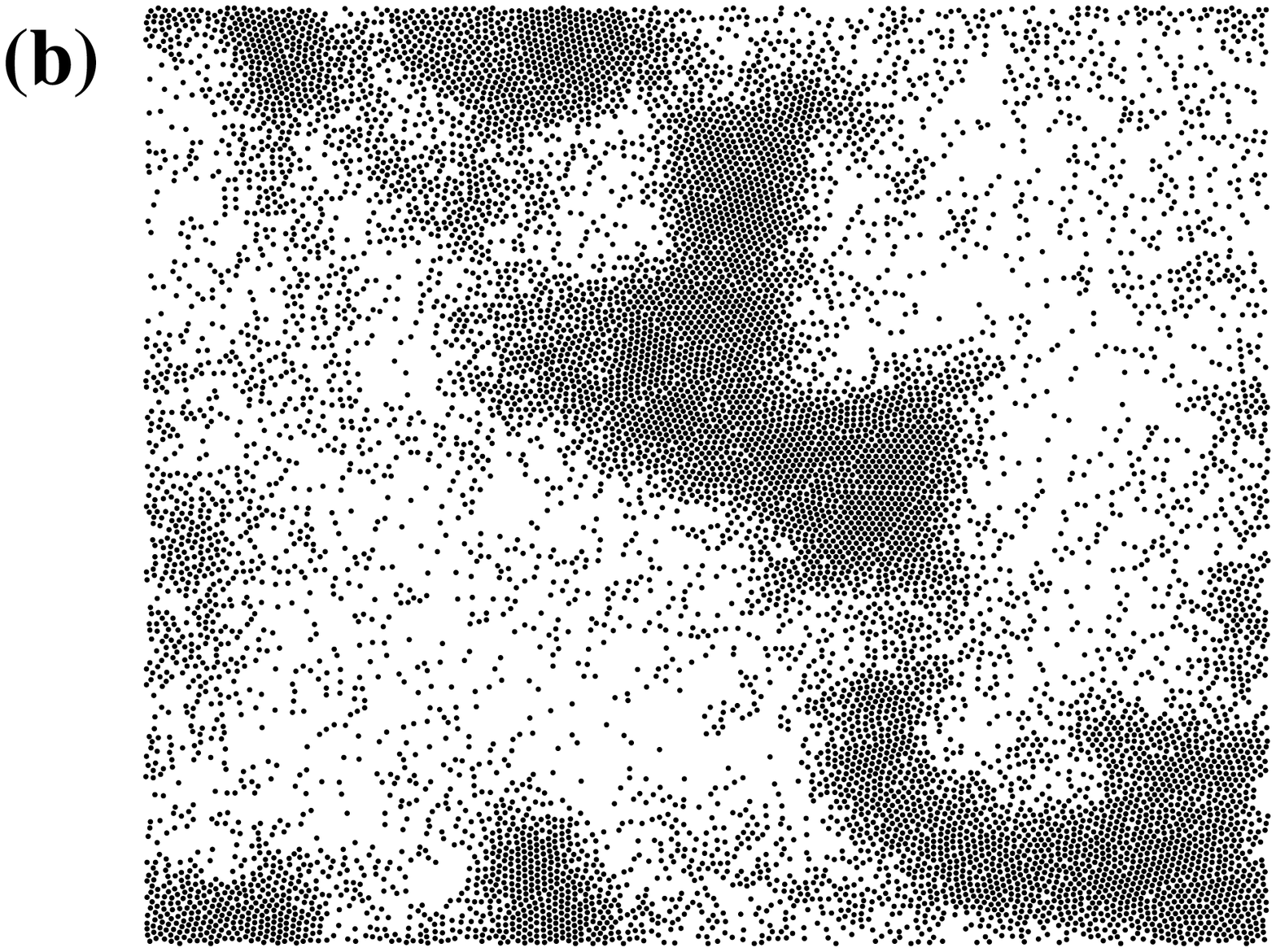,height=5.0cm,angle=-90} 
\end{center}
\protect{\caption{Snapshots of the particle distribution in the
steady state for a system of $N=11025$ particles, $\delta=1$, $\nu=0.34$,
and $H_{\rm dr}=1.0$\,s$^{-1}$, with $r=0.97$ (a) and $r=0.6$ (b).}
\label{fig:cluster}
}
\end{figure}
In Fig.\ \ref{fig:cluster} snapshots of the system's steady
state are shown for $r=0.97$ (a) and $r=0.6$ (b), with $\delta=1$.
Different regimes are observed: A homogeneous state exists for
very weak dissipation ($r=0.999$, not shown here), whereas dense, 
persistent clusters with a crystalline structure, domain boundaries, and 
vacancies are found for higher dissipation ($r=0.97$). 
The region between these clusters appears rather homogeneous
and dilute (gas-like), very similar
to the structures observed in experiments \cite{olafsen98}. 
For higher dissipation ($r=0.6$) the clusters appear less 
symmetric and are smaller.  

Note that the clustering in the case $\delta=1$ is 
qualitatively different from the case of homogeneous driving, 
and it appears already for quite high values of $r$ 
(see Fig. \ref{fig:cluster}). Homogeneous driving, in fact, 
leads to transient clusters, i.e.~they appear and disappear
continuously, while in the experiments \cite{olafsen98} and
in the case of multiplicative driving, the individual clusters are
in equilibrium with a gas phase and are stable for rather long times. 
Simulations with negative $\delta$, (we used $\delta=-0.5$ and 
$\delta=-0.25$) give a behavior qualitatively similar to the 
case of homogeneous driving ($\delta=0$).  For $\delta=1$, the 
particles inside the cluster, with rather small relative velocities,
are much less agitated than the particles in the surrounding gas
phase, so that a cluster is stable. For $\delta \le 0$, the
particles in the cluster are driven comparatively strong, what leads to
less stable, dynamic clusters.

\begin{figure}[htb]
\begin{center}
{\epsfig{file=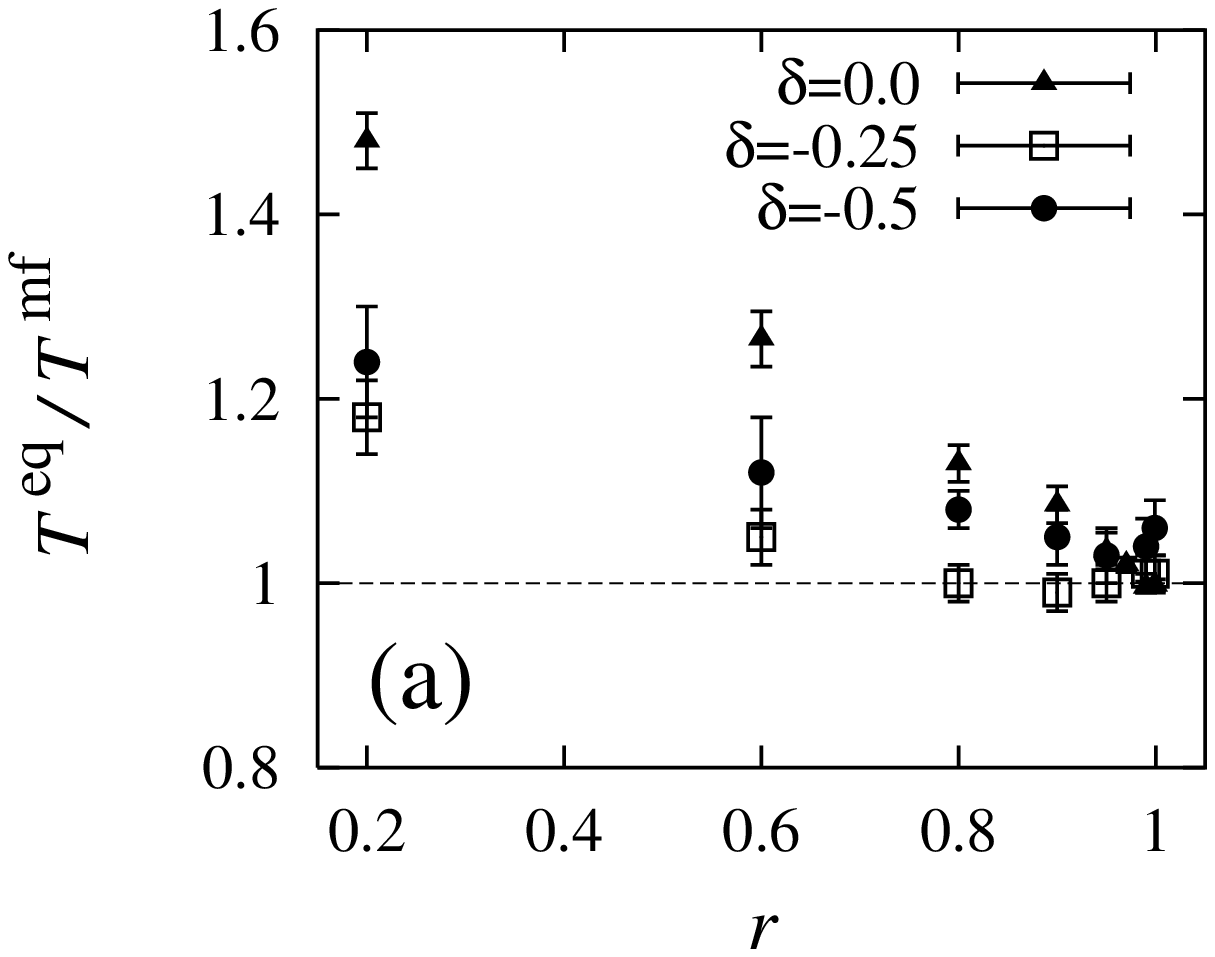, width=3.7cm,angle=-90}}
{\epsfig{file=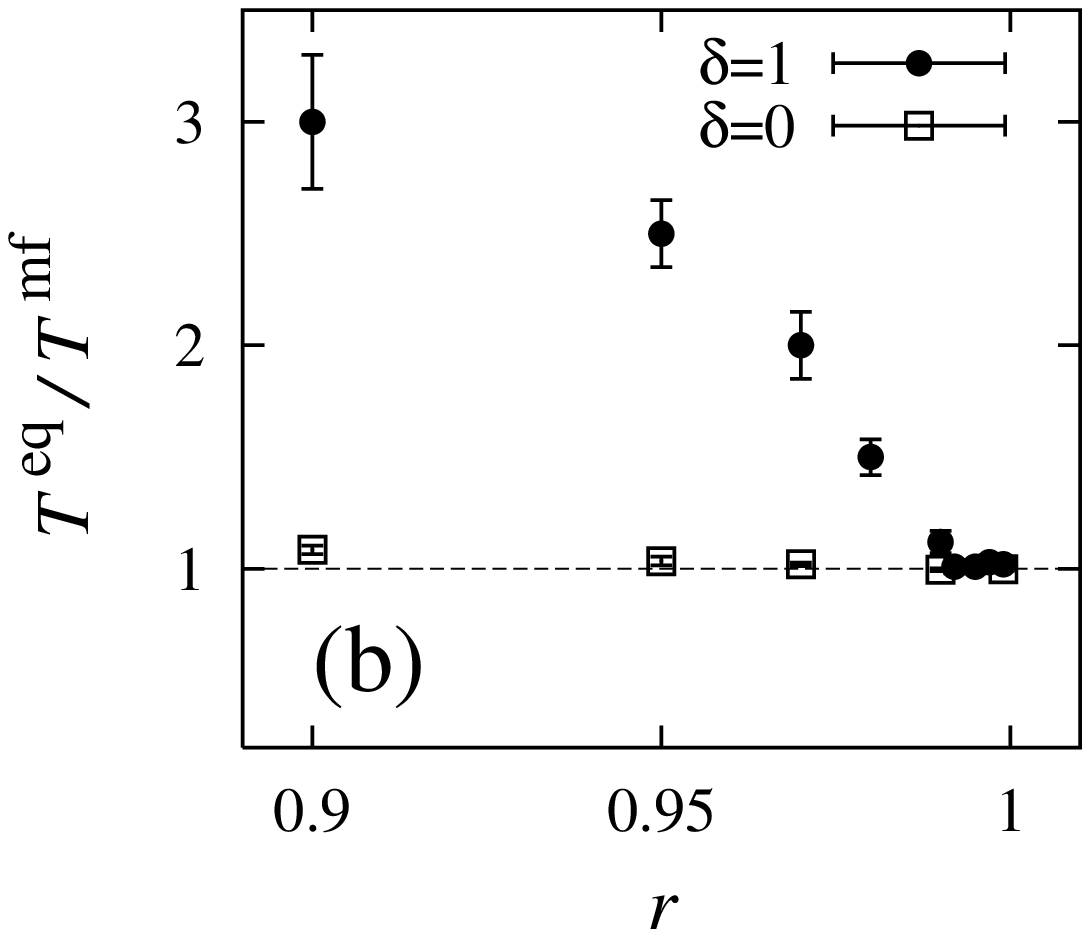, width=3.7cm,angle=-90}}\\
\end{center}
\protect{\caption{Rescaled translational temperature $T^{\rm eq}/T^{\rm mf}$
plotted against the restitution coefficient $r$ for $N=11025$
and different values of $\delta$ as given in the insets. Note the different
axis scaling in (a) and (b).}
\label{fig:vary_r}}
\end{figure}

In Fig.\ \ref{fig:vary_r} we plot the ratio between the numerical 
results for long times $T_{}^{\rm eq}$ and the theoretical equilibrium 
temperatures $T_{}^{\rm mf}$ as function of $r$ for different $\delta$. 
The agreement of the simulations with the MF prediction 
is optimal for $r \rightarrow 1$. For $\delta<0$, the range 
of agreement extends to much smaller $r$ values, i.e.~to stronger 
dissipation, as in the case of $\delta=1$ and even in the case $\delta=0$.
We have also performed simulations with
$\delta=-1$, and found that the driving is very singular in the low
velocity limit, i.e.~an excessive amount of energy is given to slow
particles, creating an inhomogeneous state.

\begin{figure}[htb]
\begin{center}
\epsfig{file=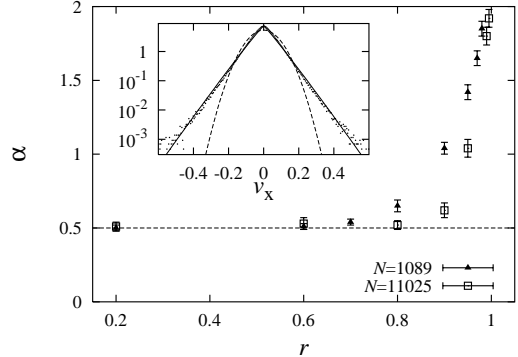,height=7.0cm,angle=-90}\\
\end{center}
\protect{\caption{Plot of the exponent $\alpha$ 
as a function of $r$, for system size $N=1089$
(solid triangles) and $N=11025$ (open squares). 
In the inset,
the dots give the averaged, normalized velocity distribution from 
simulations with $N=1089$ and $r=0.9$, the solid line is a fit 
according to $f_\alpha(v_x)$ with $f_0=7.6$, $B=18.9$ and $\alpha=1.05(9)$,
and the dashed line is a Gaussian with the same standard deviation as 
the data. }
\label{fig:vdist} }
\end{figure}

In the following, we focus on the steady state velocity distribution.
Therefore, we performed a large series of 
simulations for different values of $r$, with $\nu=0.34$,
$\delta=1$, and both $N=1089$ and $N=11025$. For each value of $r$ 
we performed about $10^3$ simulations for $N=1089$ and about $2\times 10^2$ 
simulations for $N=11025$, each with different initial configurations. 
The distribution of velocity is symmetric and isotropic
so that we present only the data for the $x$ component. 
For data analysis, a Gaussian with the width obtained from the numerical 
simulations can be superposed to the velocity distribution function
$f(v_x)$ in order to visualize the differences. Furthermore, 
as a more quantitative approach, a three parameter fit function
$f_\alpha(v_x)=f_0 \exp(-B|v_x-\langle{v_x}\rangle|^{\alpha})$ 
is used to estimate the exponent $\alpha$ of the tail of 
the distribution. The results are reported in Fig.\ \ref{fig:vdist}, 
where a representative velocity distribution is shown in the inset.
The main outcome of our simulations is that the velocity distribution 
is not Gaussian for high inelasticity and that the exponent $\alpha$ in
the stretched exponential depends on the normal restitution $r$.
The non-Gaussian behavior is limited to the tail for large $r$, but 
seems to extend over the whole range of velocities if $r$ is small enough. 
In Fig.\ \ref{fig:vdist}, the exponent $\alpha$ varies
continuously between $0.5$ and $2$ and, in particular, 
for $N=1089$ and $r=0.9$ we get $\alpha=1.05(9)$, while 
for $N=11025$ and $r=0.95$ we observe $\alpha=1.02(6)$.
Such an almost exponential velocity distribution has recently been 
observed in experiments on a vibrated monolayer of spheres, together 
with the cluster structure discussed above 
\cite{olafsen98}. 
The size dependency of our results, seems reasonable, since also in 
freely cooling systems the value of $r$ below which the inhomogeneity and 
the clustering become important grows with the system size 
\cite{goldhirsch93,luding98d,cafiero99}. 
Simulations for $\delta=-0.25$ and $-0.5$, show that the velocity 
distribution is very near to a Gaussian even for strong
dissipation. 

In order to study the shape of the velocity distribution function 
analytically, we extend the approach used by van Noije et al.\ 
\cite{n98} in the special case $\delta=0$ to arbitrary $\delta$.
The Enskog-Boltzmann equation for the freely cooling gas of spheres,
with the standard collision integral $I(f,f)$, has to be extended by
the multiplicative driving term proportional to the particle velocity.
The Enskog-Boltzmann equation, see Eq.\ (26) in \cite{n98}, 
corrected by a Fokker-Planck diffusion term \cite{footnote1}, is
\begin{equation}
\frac{\partial}{\partial t} f({\vec v}_1,t) = g(2a) I(f,f) 
   + \frac{H_{\rm dr}}{2m} \, \frac{\partial^2}{\partial{\vec v}_1^2} 
     [v_1^{2\delta} f({\vec v}_1,t)] ~.
\label{eq:fp}
\end{equation}
Considering the stationary limit $\partial f({\vec v}_1,t)/\partial t=0$,
and introducing a scaled distribution function,
with the dimensionless velocity $c=v/v_0$ and the mean thermal 
velocity $v_0$, one obtains a dimensionless evolution equation for
the scaled $f({\vec v})$, as in \cite{n98}. Multiplied
by $c_1^p$ and integrated over ${\vec c}_1$, this leads to
a set of equations which couple the moment 
$\langle{c^{p-2+2 \delta}}\rangle$ to 
the $p$-th moment $\mu_p$ of the collision term.
In the special case $\delta=1$ and $p=2$, one obtains
$v_0={H_{\rm dr}}/{(g(\nu) \mu_2 a n)}$.
For $\delta \neq 1$ and $\delta \ne 0$ the MF thermal 
velocity can be obtained by assuming
$\langle c^{2\delta}\rangle=[\langle c^{2} 
\rangle]^{\delta}=v_0^{2\delta}$ \cite{footnote3}.

For high $c$, one has $\tilde{I}(\tilde{f},\tilde{f})
\approx -\beta_1 c_1 \tilde{f}(c_1)$, 
with $\beta_1=\pi^{1/2}$ in 2D and the equation for $f(c)$ becomes
\begin{equation}
- \beta_1 c \tilde{f}(c) + \frac{\mu_2}{2d} \left(\frac{{\rm
d}^2}{{\rm d} c^2} + \frac{d-1}{c} \frac{\rm
d}{{\rm d} c}\right) [c^{2\delta}\tilde{f} (c)]=0 ~,
\end{equation}
Inserting a solution of the form $\tilde{f}(c)\propto\exp(-B c^\alpha)$, 
we obtain the large $c$ solution $\alpha=\frac{3-2\delta}{2}$ and
$B=\frac{2}{3-2\delta} \sqrt{{2d\beta_1}/{\mu_2}}$. 
In particular, for $\delta=1$ we find $\alpha=1/2$
what corresponds to the exponent we found in 
the low $r$ limit, see Fig.\ \ref{fig:vdist}. Our explanation is that
there is a crossover velocity $c^*$ 
above which the tail with exponent $1/2$ appears, which decreases as 
$r$ decreases. Such a tail 
is not observed numerically for high $r$ values, since $c^*$ is outside 
the velocity range we can examine. Simulations with $r=0.2$, $N=1089$,
and $\delta=-0.5$, $-0.25$, and $0.5$ give respectively 
$\alpha=2.10(9)$, $1.90(9)$, and $1.10(7)$, quite near to the values 
$\alpha=2$, $7/4$, and $1$, given by the large $c$ analysis.
Since the driving depends on the velocity, it acts on all the velocity 
scales, affecting the shape of $f(v)$ in a more 
complex way than in the case of homogeneous driving. Moreover, 
when inhomogeneities are present, 
the driving will depend on the local thermal velocity $v_0(\vec{x})$ 
{\em and} on the local density $n(\vec{x})$, while 
homogeneous driving is {\em identical everywhere}. 
Since inhomogeneities are generated by the dissipation, 
a {\em feed-back} between driving and dissipation could be at 
the origin of the non-universal behavior of $f(v)$. 

Future research includes a MF and clustering study for a 
wider parameter range, including the case of rough particles. 
Both a more detailed analysis of $f(v)$ \cite{cafiero99} and a 
quantitative analysis of the clusters \cite{luding99} are in progress. 
A microscopic justification for the 
multiplicative driving is still lacking and requires more detailed
experimental or three-dimensional simulation \cite{nie99} studies. 
However, we suggest  
a possible experimental check of our ideas. From MF theory the 
scaling of the equilibrium temperature of the vertically vibrated gas 
with the area fraction $\nu$ is given by 
$T^{\rm mf} \propto [\nu g(\nu)]^{-\frac{2}{3-2\delta}}$. 
Experimental measurement of the equilibrium temperature of 
the vertically vibrated monolayer, for different values of $\nu$, 
could allow to estimate 
the value of $\delta$ and to verify the hypothesis of 
a multiplicative effective driving. \\

R. C. and  H. J. H. acknowledge financial support 
under the European network project FMRXCT980183; S. L. 
and H. J. H. acknowledge funding from the Deutsche Forschungsgemeinschaft
(DFG).  We thank M. A. Mu\~noz, E. Trizac, and J. S. Urbach for helpful
discussions.


\begin{thebibliography}{10}

\bibitem{haff83}
P.~K. Haff, J. Fluid Mech. {\bf 134},  401  (1983).

\bibitem{herrmann98}
{\em Physics of dry granular media - NATO ASI Series E 350}, edited by H.~J.
  Herrmann, J.-P. Hovi, and S. Luding (Kluwer Academic Publishers, Dordrecht,
  1998).

\bibitem{goldhirsch93}
I. Goldhirsch and G. Zanetti, Phys. Rev. Lett. {\bf 70},  1619  (1993).
S. McNamara and W.~R. Young, Phys. Rev. E {\bf 53},  5089  (1996).

\bibitem{noije98}
T.~P.~C. van Noije and M.~H. Ernst, Physica A {\bf 251},  266  (1998).

\bibitem{n98}
T.~P.~C. van Noije, M.~H. Ernst, E. Trizac, and I. Pagonabarraga,
  Phys. Rev. E {\bf 59}, 4326 (1998);
T.~P.~C. van Noije and M.~H. Ernst, cond-mat/9803042.

\bibitem{luding98d}
S. Luding, M. Huthmann, S. McNamara, and A. Zippelius, Phys. Rev. E {\bf 58},
  3416  (1998).
S. Luding and S. McNamara, Granular Matter {\bf 1},  113  (1998),
  cond-mat/9810009.


\bibitem{cafiero99}
R. Cafiero, S. Luding, and H.~J. Herrmann, (unpublished).

\bibitem{huthmann98}
M. Huthmann and A. Zippelius, Phys. Rev. E {\bf 56},  6275  (1998).


\bibitem{kudrolli97}
A. Kudrolli and J.~P. Gollub,  in {\em Powders \& Grains 97} (Balkema,
  Rotterdam, 1997), p.\ 535.

\bibitem{olafsen98}
J.~S. Olafsen and J.~S. Urbach, Phys. Rev. Lett. {\bf 81},  4369  (1998),
  cond-mat/9807148.
J.~S. Olafsen and J.~S. Urbach, cond-mat/9905173 (unpublished).

\bibitem{losert98}
W. Losert, D.~G.~W. Cooper, and J.~P. Gollub, cond-mat/9812089 (unpublished).

\bibitem{nie99}
X.\,Nie, E.\,Ben-Naim, and S.\,Y.\,Chen, cond-mat/9910371.

\bibitem{duparcmeur95}
Y.~L. Duparcmeur, H.~J. Herrmann, and J.~P. Troadec, J. Phys. I France {\bf 5},
   1119  (1995).


\bibitem{puglisi98}
A. Puglisi, V. Loreto, U.~M.~B. Marconi, and A. Vulpiani, Phys. Rev. E {\bf
  59},  5582  (1999).

\bibitem{bizon99}
C. Bizon, M.~D. Shattuck, J.~B. Swift, and H.~L. Swinney, cond-mat/9904135
  (unpublished).

\bibitem{jenkins85b}
J.~T. Jenkins and M.~W. Richman, Phys. of Fluids {\bf 28},  3485  (1985).

\bibitem{footnote1}
{In the homogeneous situation ($\delta=0$), one can identify our energy input
rate with the term $m \xi_0^2$ in Ref.\ \cite{noije98}}.

\bibitem{footnote2}
{For large $\delta$ the driving rate can grow faster than the 
dissipation rate. $\delta=3/2$ is a singular limit of the MF theory, but the
instability of the homogeneous state appears already below this value (e.g.~at
$\delta=1$)}.

\bibitem{footnote3}
{The step is certainly true for $\delta=0$ and $\delta=1$.
For real values of $\delta$, we see {\em a posteriori} 
that the step leads to the correct MF theory, but we have no 
clear justification for it}.

\bibitem{luding99}
S. Luding, H.~J. Herrmann, Chaos {\bf 9}(3), 673 (1999).

\end{thebibliography}

\end{document}